\documentclass[aps,prb,reprint,superscriptaddress]{revtex4-1}

\usepackage{graphicx}
\usepackage{color}
\usepackage{amsmath}
\usepackage{bbm}

\renewcommand{\vec}{\mathbf}
\begin{document}

\title{Tuning Rashba spin-orbit coupling in homogeneous semiconductor nanowires}
\author{Pawe{\l} W\'ojcik}
\email{pawel.wojcik@fis.agh.edu.pl}
\affiliation{AGH University of Science and Technology, Faculty of
Physics and Applied Computer Science, 30-059 Krakow, Poland
Al. Mickiewicza 30, 30-059 Krakow, Poland}
\author{Andrea Bertoni}
\email{andrea.bertoni@nano.cnr.it}
\affiliation{CNR-NANO S3, Institute for Nanoscience, Via Campi 213/a, 41125 Modena, Italy}
\author{Guido Goldoni}
\email{guido.goldoni@unimore.it}
\affiliation{Department of Physics, Informatics and Mathematics, University od Modena and Reggio Emilia, Italy}

\date{\today}

\begin{abstract}
We use $\vec{k}\cdot\vec{p}$ theory to estimate the Rashba spin-orbit coupling (SOC) in large semiconductor nanowires. 
We specifically investigate GaAs- and InSb-based devices with different gate configurations to control symmetry and localization of the electron charge density.
We explore gate-controlled SOC for wires of different size and doping, and we show that in high carrier density SOC has a non-linear electric field susceptibility, due to large reshaping of the quantum states.
We analyze recent experiments with InSb nanowires in light of our calculations.
Good agreement is found with SOC coefficients reported in Phys.~Rev.~B \textbf{91}, 201413(R) (2015), but not with the much larger values reported in Nat~Commun., \textbf{8}, 478 (2017).
We discuss  possible origins of this discrepancy.
\end{abstract}


\maketitle

\section{Introduction}

Semiconductor nanowires (NWs) are attracting increasing interest for (ultra-fast) electronic and optoelectronic applications, including single-photon sources,\cite{Reimer2011} field effect transistors,\cite{Xiang2006} photovoltaic cells,\cite{Czaban2009} thermoelectric devices,\cite{Erlingsson2017} lasers\cite{Stettner2016,Holmes2014} and programmable circuits.\cite{Yan2011} Recently, special attention raised for spintronic applications\cite{Fabian2007,RossellaNN2014,Wojcik2014} and topological quantum computing.\cite{Nayak2008} 
Due to strong spin-orbit coupling (SOC) in InSb- or InAs-based NWs, an \emph{helical} gap has been observed if a finite magnetic field is applied orthogonal to the SOC effective field, $B_{SOC}$.\cite{Pershin2004,Streda2003,Quay2010,Kammhuber2017,Heedt2017} In this 1D state, carriers with opposite momentum have opposite spin. 
In combination with the proximity-induced superconductivity,\cite{Krogstrup2015,Chang2015} it imitates spin-less $p$-wave superconductor (Kitaev chain),\cite{Kitaev2003} making the strongly spin-orbit coupled InSb and InAs NWs possible host materials for topologically protected quantum computing based on Majorana zero modes.\cite{Mourik2012,Alicea2012,Sau2012,Albrecht2016,Manolescu2017} 

SOC is a relativistic effect where a part of the electric field is seen as an effective magnetic field in the charged particle rest frame. In semiconductor crystals, the electric field may arise from a symmetry breaking that is either intrinsic, i.e., related to the crystallographic structure of the material (Dresselhaus SOC),\cite{Dresselhaus} or induced by the overall asymmetry of the confinement potential due to an electrostatic field, due to, e.g., compositional profiles, strain, or external gates (Rashba SOC).\cite{Rashba} Typically, SOC is the combination of both components,\cite{ClimentePRB2007} but zincblende NWs grown along [111] posses inversion symmetry, and the Dresselhaus contribution vanishes. This NW direction is the one used in experiments exploring the existence and nature of Majorana bound states.\cite{Mourik2012,Albrecht2016} Therefore, we shall consider only the Rashba SOC throughout the paper.

A critical issue in this context is to engineer devices with strong SOC, as the ratio of the spin-orbit energy relative to the Zeeman energy determines the magnitude of topological energy gap protecting zero-energy Majorana modes.\cite{Alicea2012} Recent studies of 2D InSb wires and planar InSb heterostructures show a SOC constant $\alpha_R=3$~meVnm.\cite{Kallaher2010,Kallaher2010b} Larger values were reported for quantum dots gated in InSb NWs, $\alpha_R=16-22$~meVnm,\cite{Nilsson2009,Nadjperge2012} which likely includes a contribution from the local electric fields of the confining gates.

The standard method to extract the SOC in semiconductor NWs is by magnetoconductance measurements in low magnetic fields, exploiting the negative magnetoresistance due to weak anti-localization.\cite{Bergmann1984,Kettemann2007}
Recently, this technique was used to extract SO strength in InSb NWs demonstrating very large values of the SOC constant, $\alpha_R = 50-100$~meVnm.\cite{vanWeperen2015}
Unexpectedly, a much higher value of $\alpha_R$ was reported in Ref.~\onlinecite{Kammhuber2017} where the authors used the conductance measurement technique.
The measure of the conductance through the helical state and comparison of the data to the theoretical model gives a spin-orbit energy $E_{SO}=6.5$~meV, which corresponds to $\alpha_R=270$~meVnm, the highest value reported so far for semiconductor NWs.

The determination of SOC strength in semiconductor NWs still remains an open issue, with different measurement techniques leading to values of $\alpha_R$ differing by almost one order of magnitude. It should be noted that for typical samples, with diameters in the tens of nm range, the symmetry and localization of the quantum states is a delicate balance between different energy scales and it is strongly influenced by external fields.\cite{MorkotterNL2015,RoyoPRB2013,JadczakNL2014,ManolescuPRB2016} 
On the other hand, theoretical investigations of SOC so far\cite{Kokurin2015,Kokurin2014} only rely on simple models which do not capture the complexity of the quantum states whose symmetry underlies the Rashba contribution to SOC nor its tunability by an electric field, which is the goal of the present study.

In this paper we evaluate the SOC strength on the basis of a $\vec{k}\cdot\vec{p}$ theory using self-consistent quantum states which take into account the realistic geometry of large, doped NWs. Our analysis includes external metallic gates and dielectric layers of typical NW-based devices. We evaluated the SOC coefficients as a function of NW size and gate configuration. We find that the strong interplay between external fields and the localization of quantum states results in a strong non-linear electric field susceptibility for SOC in the high carrier density regime. We analyze recent experiments with InSb NWs in light of our calculations. Good agreement is found with SOC reported in Phys.~Rev.~B \textbf{91}, 201413(R) (2015), but not with the much larger values measured in Nat~Commun., \textbf{8}, 478 (2017), and we discuss  possible origins of this discrepancy.

The paper is organized as follows. In Sec.~\ref{sec:TheoreticalModel} we obtain SOC coefficients from the $\vec{k}\cdot\vec{p}$ theory. The effective Hamiltonian which determines the quantum states is devised in Sec.~\ref{sec:EffectiveHamiltonian}, while SOC coefficients  in terms of the envelope functions of the structure are derived in Sec.~\ref{sec:SOC}. In Sec.~\ref{sec:Results} we apply our methodology to GaAs-based (Sec.~\ref{sec:GaAs}) and InSb-based (Sec.~\ref{sec:InSb}) devices, discussing recent experiments in the latter case. Summary of our investigation is drawn in Sec.~\ref{sec:summary}.

\section{Theoretical model}
\label{sec:TheoreticalModel}

\begin{figure}[!h]
	\begin{center}
		\includegraphics[scale=0.4]{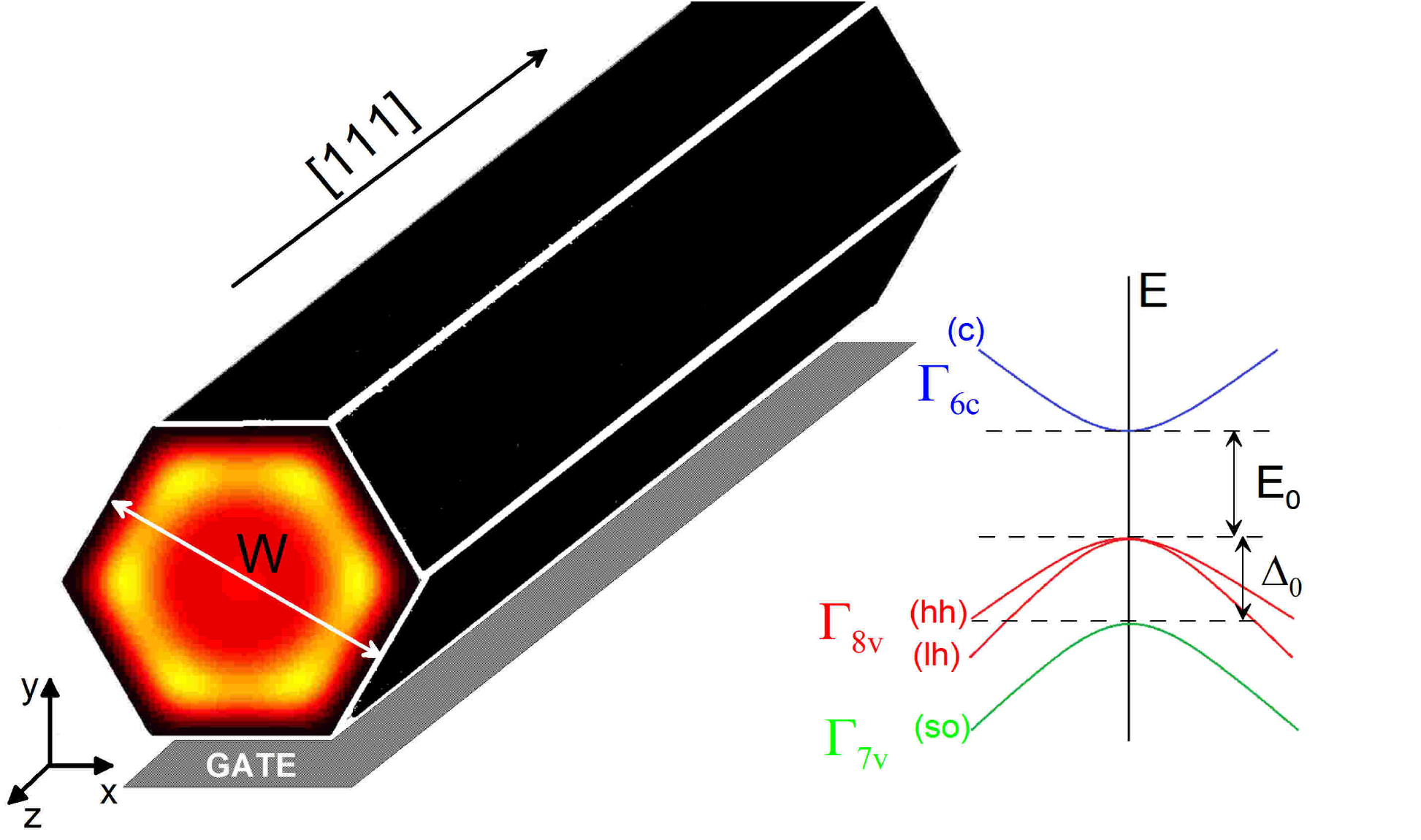}
		\caption{Schematic illustration of a NW with a bottom gate. A typical electron gas distribution is shown inside (yellow). A schematic of the semiconductor band structure used within the Kane model is shown. Symbols indicate conduction (c), heavy-hole (hh), light-hole (lh), and split-off (so) bands with corresponding group-theoretical classification of zone center states.}
		\label{fig1}
	\end{center}
\end{figure}

Our target systems are NWs with hexagonal cross-section, grown in the [111] direction, see Fig.~\ref{fig1}. In these systems, quantum states are determined by several sample parameters, including geometry, Fermi energy, external fields, etc. Below we use the $8 \times 8$ Kane model to derive the (Rashba) SOC constants in terms of a realistic description of the quantum states. This allows for quantitative predictions of SOC constants as a function of the gate voltages and geometrical parameters in different regimes and gate configurations. 

\subsection{Effective Hamiltonian for SOC of conduction electrons}
\label{sec:EffectiveHamiltonian}

Formally, our target system has translational invariance along $z$. Each component of the envelope function (one for each total angular momentum component) can be developed in a set of subbands $\psi_n(x,y)$, the coefficients of the linear combination being determined by the Kane Hamiltonian. Here, $n$ is a subband index and $(x,y)$ are the space directions in a plane sectioning the NW. Since NWs in our calculations are quite large (with diameters $\sim 10^2$ nm), it is usually necessary to include a large number of subbands in calculations.

The $8 \times 8$ Kane Hamiltonian reads\cite{Fabian2007} 
\begin{equation}
\label{eq:KH}
H_{8\times 8}=\left ( 
\begin{array}{cc}
H_c & H_{cv} \\
H^{\dagger}_{cv} & H_v
\end{array}
\right ),
\end{equation}
where $H_c$ is the $2\times 2$ diagonal matrix related to the conduction band ($\Gamma_{6c}$ at the $\Gamma$ point of Brillouin zone, see Fig.~\ref{fig1}) while $H_v$ is the $6\times 6$ diagonal matrix corresponding to the valence bands ($\Gamma _{8v}$, $\Gamma _{7v}$)
\begin{eqnarray}
H_c&=&H_{\Gamma _6}(x,y)\mathbf{1}_{2\times2},  \\
H_v&=&H_{\Gamma _8}(x,y)\mathbf{1}_{4\times4} \oplus H_{\Gamma _7}(x,y)\mathbf{1}_{2\times2}.
\end{eqnarray}
In the above expressions
\begin{eqnarray}
H_{\Gamma _6}(x,y)&=& -\frac{\hbar ^2}{2m_0} \nabla _{2D}^2 + \frac{\hbar ^2 k_z^2}{2m_0} +
E_c+ V(x,y), \\
H_{\Gamma _8}(x,y)&=& E_c+ V(x,y)-E_0, \\
H_{\Gamma _7}(x,y)&=& E_c+ V(x,y)-E_0 - \Delta _0,
\end{eqnarray}
where $\nabla _{2D}=(\frac{\partial}{\partial x},\frac{\partial}{\partial y})$, $m_0$ is the free electron mass, $E_c$ is the energy of the conduction
band edge, $E_0$ is the energy gap, $\Delta_0$ is the split-off band gap and  $V(x,y)$ is the potential energy. In doped systems, the potential $V(x,y)$  consists of the sum of the Hartee potential energy generated by the electron gas and ionized dopants, and any electrical potential induced by gates attached to the NW, $V(x,y)=V_H(x,y)+V_{gate}(x,y)$.  We adopt the hard wall boundary conditions at the surface of NWs. 

The off-diagonal matrix $H_{cv}$ in (\ref{eq:KH}) reads
\begin{equation}
 H_{cv}=\left ( \begin{array}{cccccc}
         \frac{-\hat{\kappa}_+}{\sqrt{2}} & \sqrt{\frac{2}{3}}\kappa_z & \frac{\hat{\kappa}_-}{\sqrt{6}} & 0 &
\frac{-\kappa_z}{\sqrt{3}} & \frac{-\hat{\kappa}_-}{\sqrt{3}} \\
0 & \frac{-\hat{\kappa}_+}{\sqrt{6}} & \sqrt{\frac{2}{3}}\kappa_z & \frac{\hat{\kappa}_-}{\sqrt{2}} &
\frac{-\hat{\kappa}_+}{\sqrt{3}} & \frac{-\kappa_z}{\sqrt{3}} 
        \end{array}
        \right ),
\end{equation}
where $\hat{\kappa} _{\pm}=P\hat{k}_{\pm}$, $\kappa _{z}=P k_{z}$,  $\hat{k}_{\pm}=\hat{k}_x\pm i\hat{k}_y$ and $P=-i\hbar \langle S|\hat{p}_x|X
\rangle / m_0$ is the conduction-to-valence band coupling with $|S\rangle$, $|X \rangle$ being the Bloch functions at the $\Gamma$ point of Brillouin
zone.

Using the folding-down transformation, the $8\times 8$ Hamiltonian (\ref{eq:KH}) reduces into the $2\times 2$ effective Hamiltonian for the conduction band electrons 
\begin{equation}
\label{eq:Hc}
 \mathcal{H}(E)=H_c+H_{cv}(H_v-E)^{-1}H_{cv}^{\dagger}.
\end{equation}
Since $E_0$ and $\Delta _0$ are the largest energies in the system, we can expand the on- and
off-diagonal elements of the Hamiltonian (\ref{eq:Hc}) to second order in the wavevectors
\begin{eqnarray}
\label{eq:3DH}
 \mathcal{H} &=& \left [ -\frac{\hbar ^2}{2m^*} \nabla _{2D}^2 + \frac{\hbar ^2
k_z^2}{2m^*} + E_c + V(x,y) \right ] \mathbf{1}_{2\times 2} \nonumber \\
&+& (\alpha _x \sigma _x + \alpha _y \sigma _y)k_z,
\end{eqnarray}
where $\sigma _{x(y)}$ are the Pauli matrices, $m^*$ is the effective mass
\begin{equation}
 \frac{1}{m^*}=\frac{1}{m_0}+\frac{2P^2}{3\hbar ^2} \left ( \frac{2}{E_g} + \frac{1}{E_g+\Delta_g}
\right ),
\end{equation}
and $\alpha_x$, $\alpha_y$ are the SOC coefficients given by
\begin{eqnarray}
\label{eq:ax}
\alpha _x(x,y) & \approx & \frac{1}{3} P^2  \left ( \frac{1}{(E_0+\Delta_0)^2}-\frac{1}{E_0^2} \right ) \frac{\partial V(x,y)}{\partial
y}, \\
\label{eq:ay}
\alpha _y(x,y) & \approx & \frac{1}{3} P^2  \left ( \frac{1}{(E_0+\Delta_0)^2}-\frac{1}{E_0^2} \right ) \frac{\partial V(x,y)}{\partial
x}.
\end{eqnarray}

Without SOC, confinement in the $x-y$ plane of the NW leads to the formation of quasi-1D spin-degenerate subbands, with the in-plane envelope functions $\psi_n(x,y)$'s determined by the compositional and doping profiles, the field induced by the free carriers and the external gates. The 3D Hamiltonian (\ref{eq:3DH}) can be represented in the basis set $\psi_n(x,y) \exp(i k_z z)$.

The matrix elements of the spin-orbit term are given by 
\begin{equation}
\label{eq:a}
 \alpha_{x(y)}^{nm}=\int \int \psi_n(x,y) \alpha_{x(y)}(x,y) \psi_m(x,y) dx dy.
\end{equation}
These coefficients define intra- ($\alpha_{x(y)}^{nn}$) and inter-subband ($\alpha_{x(y)}^{nm}$) SOC constants which are extracted from experiments\cite{vanWeperen2015} and are estimated in Sec.~\ref{sec:Results} for several classes of material and device configurations.

\subsection{SO coupling constants calculations}
\label{sec:SOC}

To obtain the electronic states of a NW $\psi_n(x,y)$ to be used in Eq.~(\ref{eq:a}) we employ a standard envelope function approach in a single parabolic band approximation. Electron-electron interaction is treated at the mean-field level by the standard self-consistent Sch{\"o}dinger-Poisson approach. 
Assuming translational invariance along the growth axis $z$ we reduce the single-electron Hamiltonian (without SOC) to a 2D problem in the $(x,y)$ plane
\begin{equation}
\label{eq:RS2D}
\left [ -\frac{\hbar ^2}{2m^*}\nabla _{2D}^2  + E_c + V(x,y) \right ] \psi_n(x,y)=E_n \psi_n(x,y).
\end{equation}

The above eigenproblem is solved numerically by a box integration method\cite{Selberherr} on a triangular grid with hexagonal elements.\cite{RoyoPRB2014}
While this grid is symmetry compliant if the hexagonal NW is in the isotropic space, avoiding artifacts from the commonly used rectangular grid, calculations do not assume any symmetry of the quantum states. Therefore, our calculations allow to describe less symmetric situations, e.g., with external gates applied to the NW.

After solutions of Eq.(\ref{eq:RS2D}), we calculate the free electron density
\begin{equation}
 n_e(x,y)=2\sum _n \left | \psi_n(x,y) \right |^2 \sqrt{\frac{\overline{m}^* k_b T}{2\pi \hbar ^2}} \mathcal{F}_{-\frac{1}{2}} \left ( 
\frac{-E_n+\mu}{k_b T} \right ),
\end{equation}
where $\overline{m}^*$ is the effective electron mass along the NW axis, $k_B$ is the Boltzmann constant, $T$ is the temperature, $\mu$ is
the Fermi level and $\mathcal{F}_{k}=\frac{1}{\Gamma(k+1)}\int _0 ^{\infty} \frac{t^k dt}{e^{t-x}+1}$ is the complete Fermi-Dirac integration of
order $k$.

Finally, we solve the Poisson equation 
\begin{equation}
\label{eq:poisson}
 \nabla ^2_{2D} V(x,y)=-\frac{n_e(x,y)}{\epsilon _0 \epsilon},
\end{equation}
where $\epsilon$ is the dielectric constant. Equation (\ref{eq:poisson}) is solved by a box integration method on the triangular grid assuming, if not stated otherwise,  Dirichlet boundary conditions. The resulting potential $V(x,y)$ is put into Eq.~(\ref{eq:RS2D}), and the cycle is repeated until self-consistency is reached. Further details concerning the self-consistent method for hexagonal NWs can be found in Ref.~\onlinecite{Bertoni2011}.

The self-consistent potential energy profile $V(x,y)$ and the corresponding envelope  functions $\psi_n(x,y)$ are finally used to determine the SOC $\alpha^{nm}_{x(y)}$ from Eq.~(\ref{eq:a}). 

\section{Results}
\label{sec:Results}

We used the above methodology to predict SOC coefficients in different classes of materials of direct interest in NW-based spintronics. We put particular emphasis to establish the tunability of the SOC by external gates. 
Indeed, the latter strongly shape the quantum states, particularly if NWs are heavily doped, as it turns out.  We conclude this section by a qualitative comparison with the latest experiments with InSb-based NWs.

\subsection{GaAs}
\label{sec:GaAs}

\begin{figure*}[!ht]
	\begin{center}
		\includegraphics[scale=.8]{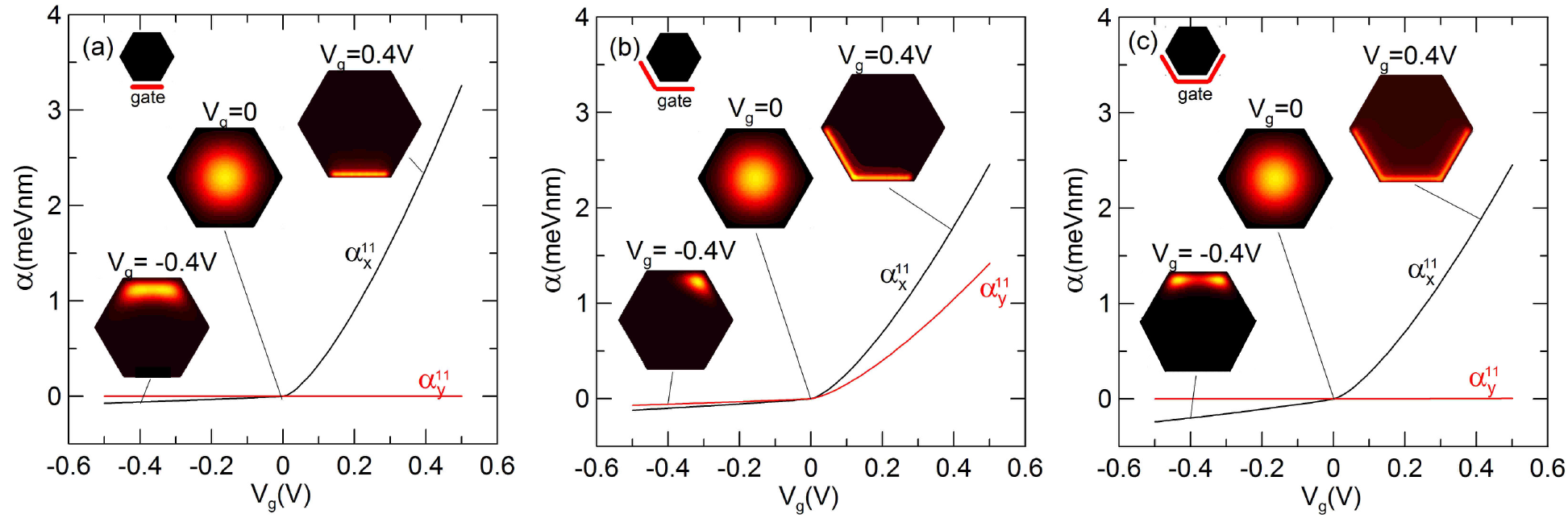}
		\caption{SOC coefficients $\alpha^{11} _x$ and $\alpha^{11} _y$ as a function of the gate voltage $V_g$ for three different gate configurations, as shown in the top-left insets. (a) bottom gate, (b) left-bottom gate and (c) left-bottom-right gate. In each panel, insets show the self-consistent electron density at gate voltages $V_g=-0.4, 0, 0.4$~V.}
		\label{fig2}
	\end{center}
\end{figure*}

GaAs is not a strong SOC material. However, it is the material of choice for transport experiments, due to its high mobility.
Recent literature reports high-mobility in doped GaAs-NWs, comparable to planar structures grown along the same crystallographic directions.\cite{FunkNL2013}
Therefore, to establish the potentiality of GaAs for spintronics, we consider GaAs homogeneous NWs with 'ideal' gate configurations, i.e., with gates directly attached to the NW.
Often, in realistic devices, a dielectric spacer layer is used in experiments. Therefore, our calculations below should be considered as an upper bound for SOC in GaAs NWs.

\begin{figure}[!ht]
	\begin{center}
		\includegraphics[scale=.4]{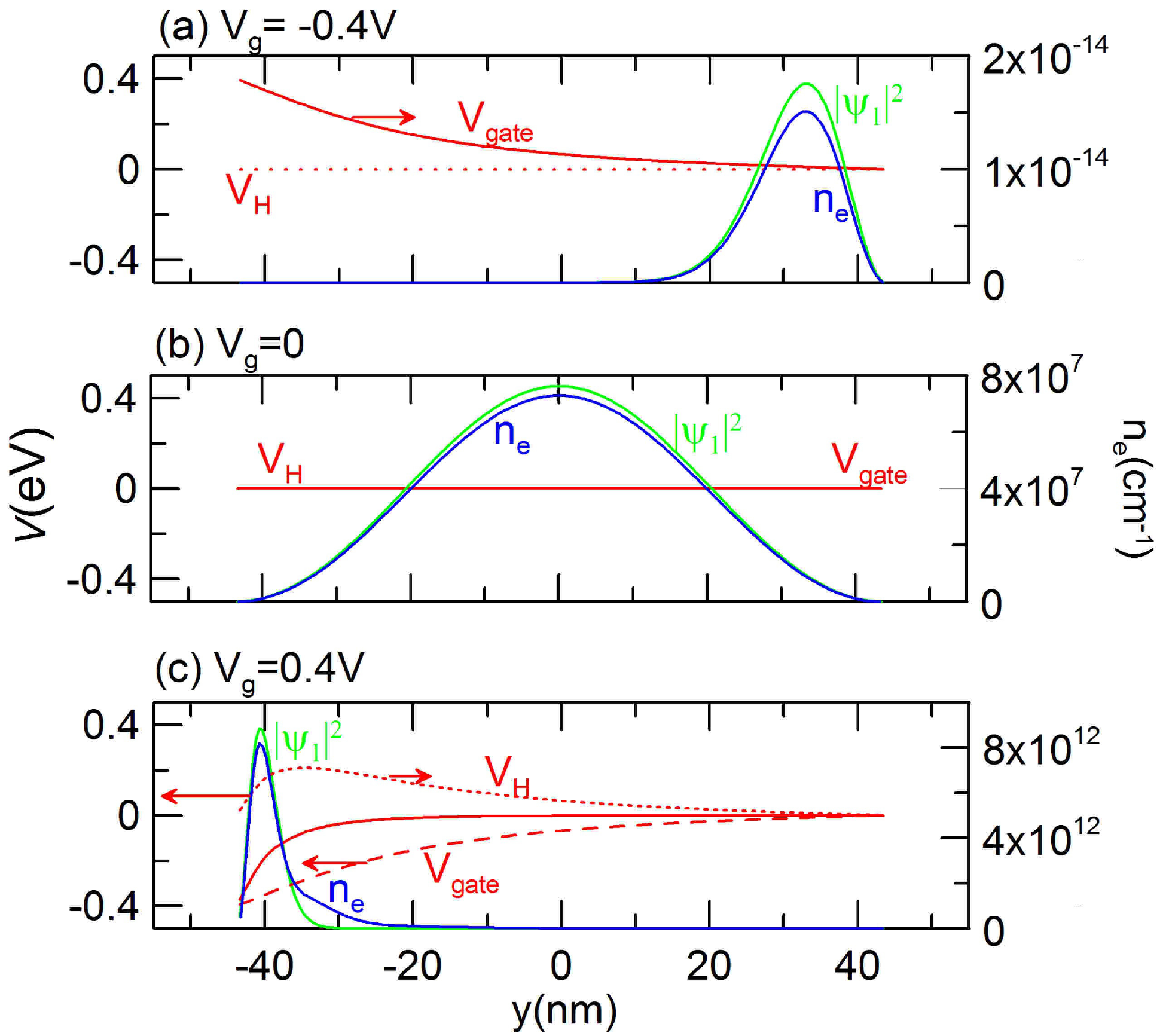}
		\caption{Cross-sections of the self-consistent potential $V$ (red line, left axis), electron density distribution $n_e$ (blue line, right axis) and $|\psi_1(x=0,y)|^2$ (green line) along the diameter in the $y$ direction for the gate voltages (a) $V_g=-0.4$~V, (b) $V_g=0$, (c) $V_g=0.4$~V. 
			The two components of the self-consistent potential $V$, namely the gate voltage $V_{gate}$ and the electron-electron interaction $V_{H}$, are shown in red dashed and dotted lines, respectively.  Arrows attached to the potential curves denote the electric field direction. Calculations correspond to the bottom-gate configuration of Fig.~\ref{fig2}(a).}
		\label{fig3}
	\end{center}
\end{figure}

The calculations have been carried out for the following material parameters:\cite{Vurgaftman2001}
$E_0=1.43$~eV, $\Delta_0=0.34$~eV, $m^*=\overline{m}^*=0.067$, $E_P=2m_0P^2 /\hbar^2 =28.8$~eV and dielectric constant $\epsilon=13.18$.
We consider a temperature $T=4.2$~K.
We assume constant chemical potential $\mu=0.85$~eV. This value ensures that only the lowest electronic state is occupied at $V_g=0$. If not stated otherwise, the calculations has been carried out for the NW width $W=87$~nm on the grid $100 \times 100$.

In Fig.~\ref{fig2} we show the SOC coefficients $\alpha^{11} _x$ and $\alpha^{11} _y$ with three different typical gate configurations, bottom gate, left-bottom gate, and U-shaped gate, as sketched in the top-left insets. 
The gates are held at a voltage $V_g$ which is swept through.
We first note that at $V_g=0$ the NW has inversion symmetry, hence $\alpha^{11} _x = \alpha^{11} _y =0$. 
As the gate voltage is switched on, $\alpha^{11} _i \ne 0$, with $i$ being the direction of the axis of symmetry broken by the field. 
So, for example, $\alpha^{11} _y =0$ in Fig.~\ref{fig2}(a),(c), but not in Fig.~\ref{fig2}(b). 
On the other hand $\alpha^{11}_x \ne 0$ in all configurations, since gates remove inversion symmetry about $x$ in all cases. 
The evolution of $\alpha^{11} _x(V_g)$ is strongly asymmetric, the strongest asymmetry being observed for the configuration with a bottom gate.
Similarly, $\alpha^{11} _y(V_g)$ is strongly asymmetric when a left-bottom gate removes inversion symmetry about both $x$ and $y$.

The behavior of the SOC coefficients results from a complex interplay between quantum confinement from the NW interfaces, the gate-induced electric field and the self-consistent field due to electron-electron interaction. 
Let us consider first the bottom gate configuration, Fig.~\ref{fig2}(a). The profile of $|\psi_1(x,y)|^2$, $n_e(x,y)$, and $V(x,y)$ are shown in Fig.~\ref{fig3} at selected gate voltages. 
To understand the impact of the individual effects on the SOC, in Fig.~\ref{fig3} the self-consistent potential $V(x=0,y)$ has been divided into two components, the one from the gate, $V_{gate}$ and the Hartree component, $V_{H}$. 
At the negative voltage $V_g=-0.4$~V the electron energy is increased by a corresponding quantity near the gate. 
Therefore, electrons are pushed away from the bottom facet of the NW, and localize near the top facets (compare with insets in Fig.~\ref{fig2}(a)). 
By the assumption of a constant chemical potential, the NW becomes highly depleted of the charge [compare the scale of the right axes in Fig.~\ref{fig3}]. 
Consequently, the electron-electron interaction is negligibly small, and the SOC, in this case, is mainly determined by the electric field coming from the gate.
Its low value is related to the localization of the ground state $\psi _1$ near the upper edge, where the gradient of the potential $\partial V(x,y) / \partial y$ is very low [Fig.~\ref{fig3}(a)]. 

The opposite situation occurs for the positive gate voltage, $V_g=0.4$~V [Fig.~\ref{fig3}(c)], at which a decrease of the conduction band by the positive voltage results in the accumulation of charge in the vicinity of the bottom gate. 
By the assumption of a constant chemical potential, the NW becomes highly doped of the charge. 
The high value of the SOC in this case is due to the electron-electron interaction which, for a high electron concentration, interplays with the gate electric field to increase SOC. 
Specifically, it almost completely compensates the gate electric field in the middle of the NW, simultaneously strengthening it near the bottom facet, where the envelope  function of the ground state localizes. 
Since this effect is stronger for the high electron concentration, the SOC coefficients significantly increases with increasing the gate voltage, in the range $V_g>0$.

\begin{figure}[!ht]
	\begin{center}
		\includegraphics[scale=.4]{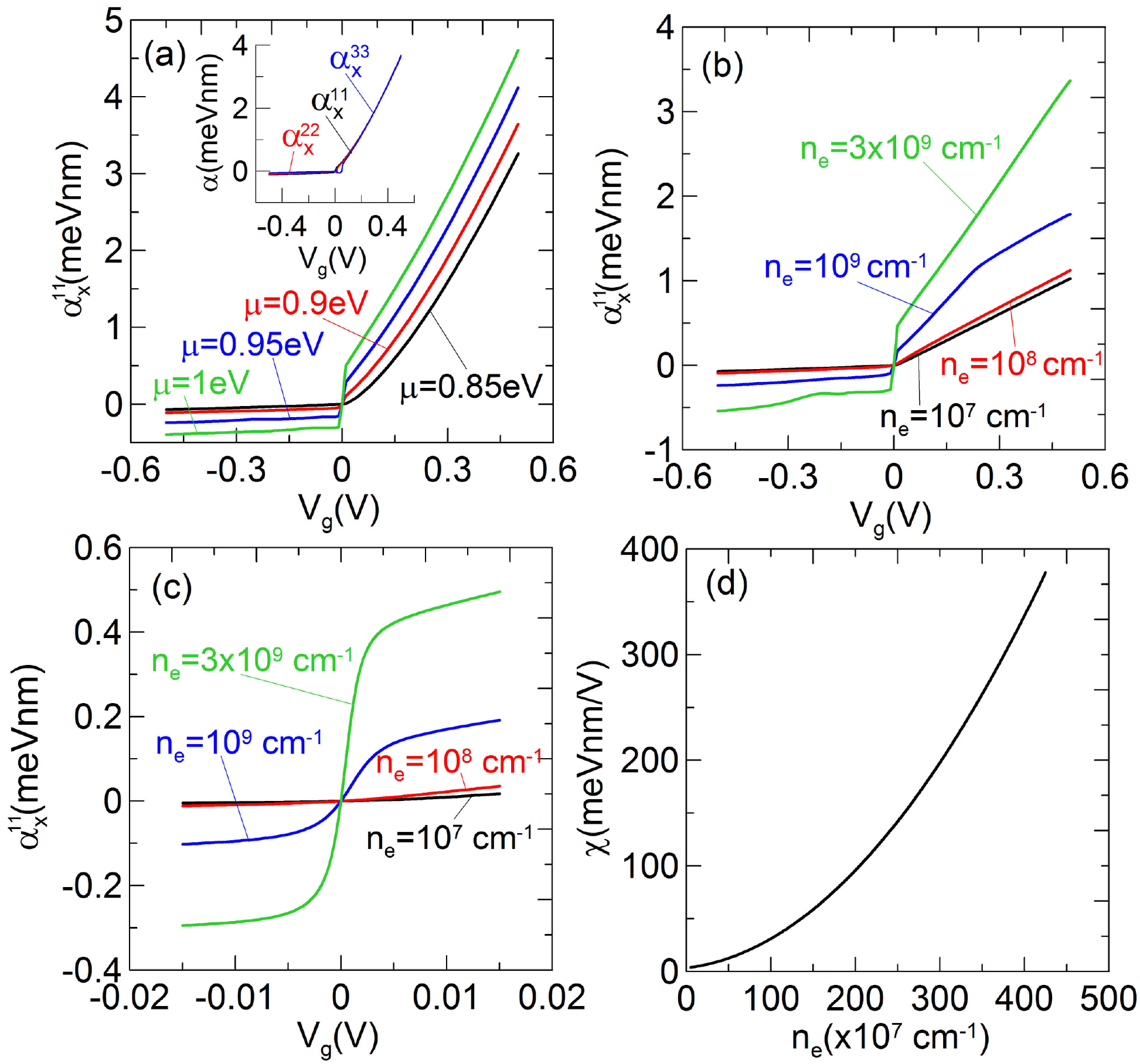}
		\caption{$\alpha^{11} _x (V_g)$ calculated under the assumption of (a) a constant chemical potential $\mu$, and (b) a constant electron density $n_e$. Inset in panel (a) shows the comparison between SOC coefficients of the three lowest subbands, calculated with constant $\mu=0.85$~eV. (c) Same as panel (b) but zooming around symmetry point $V_g=0$. (d) The SOC electric susceptibility $\alpha^{11}_x$ at
		$V_g=0$ as a function of the electron concentration $n_e$.}
		\label{fig4}
	\end{center}
\end{figure}

Figure~\ref{fig4} shows the calculated  $\alpha^{11} _x$ for constant chemical potential [Fig.~\ref{fig4}(a)] and constant electron density [Fig.~\ref{fig4}(b)].\footnote{As mentioned above, in this gate configuration $\alpha^{11} _y = 0$ by symmetry} 
$\alpha^{11} _x$ shows similar behavior in both configurations. 
Specifically, for $V_g<0$, it is almost insensitive to the gate voltage, while for $V_g>0$, it increases with $V_g$, the main difference between the two calculations being that for constant $n_e$ the behavior is almost linear, with the slope strongly dependent on the electron concentration.
Note, however, that in contrast to the $\mu$-constant model, for which the asymmetry of $\alpha^{11}_x (V_g)$ arises from the charging and discharging of the NW by the gate voltage (what determines the Coulomb interaction), for the $n_e$-constant model the asymmetry results only from the redistribution of electrons caused by the gate electric field. 

We have checked that, regardless of the electron concentration and the calculation model, the behavior of $\alpha^{nn}(V_g)$ for a few lowest subbands is almost identical. 
As an example, in the inset of panel (a) of Fig.~\ref{fig4} we show the intra-subband SOC coefficients vs.~$V_g$ for the three lowest subbands. 
These results, calculated with constant  $\mu=0.858$~eV, differ only slightly, mainly in the vicinity of $V_g=0$, where SOC is small. 
Therefore, we limit ourselves to the analysis of the intra-subband coefficient for the ground state, $\alpha^{11}_x$ throughout. 

Interestingly, for the high electron concentration (or, analogously, above a certain Fermi energy) the SOC coefficient shoots up around $V_g=0$. In Fig.~\ref{fig4}(d) we zoom in $\alpha _x^{11}(V_g)$ around $V_g=0$. The different behavior at low and high density can be understood in terms of the very different charge redistribution in the two regimes, as we discuss below. 

\begin{figure}[!ht]
	\begin{center}
		\includegraphics[scale=.4]{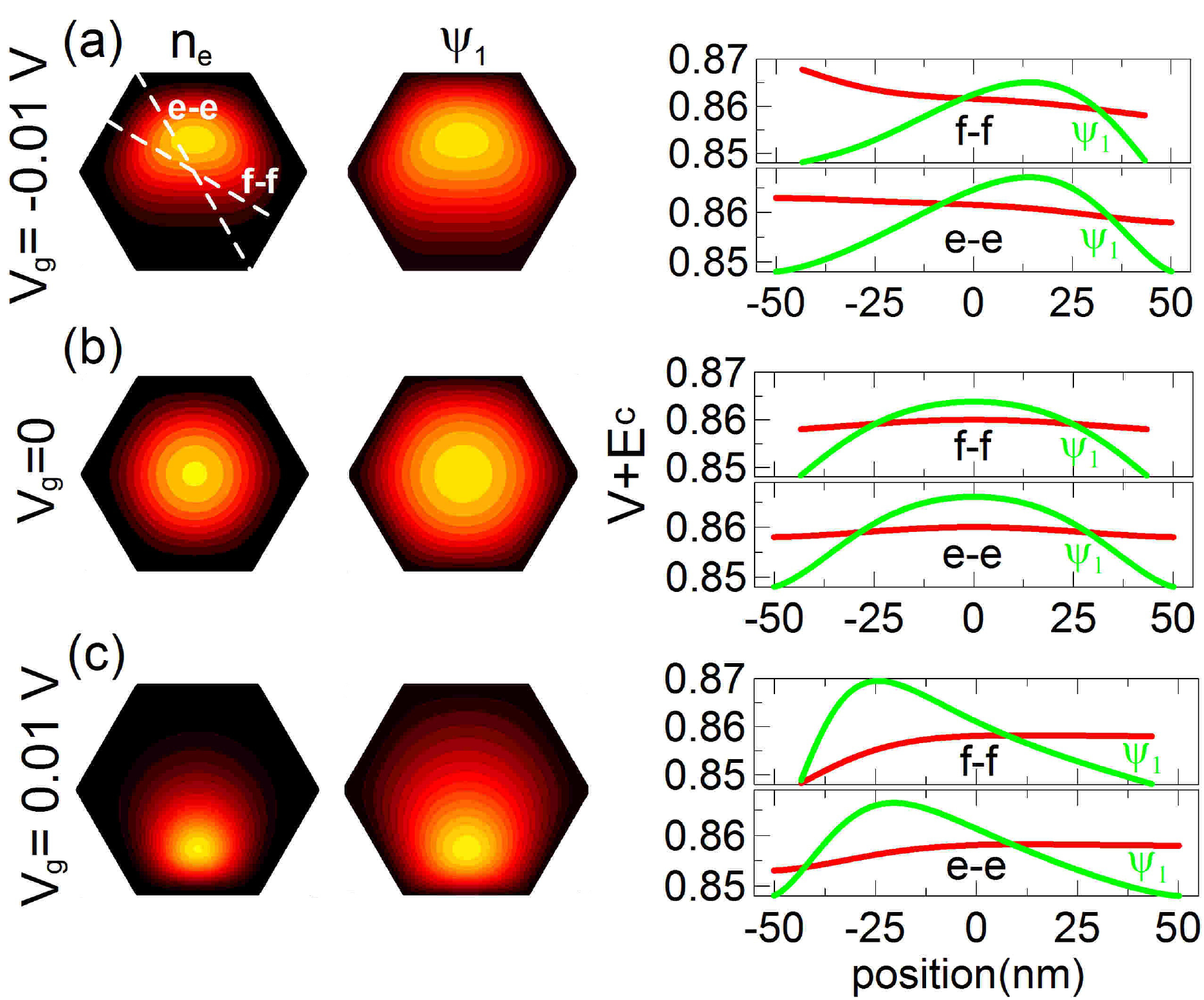}
		\caption{Left column: Maps of electron density $n_e(x,y)$ (left) and envelope  function $\psi_1(x,y)$ (right) for gate voltage (a)
			$V_g=-0.01$~V, (b) $V_g=0$ and (c) $V_g=0.01$~V. Right column: profile of self-consistent potential $V(x,y)$ (red) and the envelope  function $\psi_1(x,y)$ (green) along the facet-facet (upper) and edge-edge (lower) directions, as illustrated by the dashed lines in the top-left hexagon. Calculations performed with  $n_e=10^7$~cm$^{-1}$. }
		\label{fig5}
	\end{center}
\end{figure}

In Fig.~\ref{fig5} we show the electron density maps $n_e(x,y)$ and the envelope  function $\psi_1(x,y)$ at three distinct gate voltages around $V_g=0$, calculated for a \textit{low electron concentration}, $n_e=10^7$~cm$^{-1}$. 
The right column displays the cross-section of the self consistent potential energy $V(x,y)$ and the envelope  function $\psi_1(x,y)$ along the facet-facet vertical diameter (upper) and edge-edge diagonal diameter (lower), respectively.
In this regime, the electron-electron interaction is negligible, quantum confinement from interfaces dominates, and at $V_g=0$ the conduction band energy is nearly flat, see Fig.~\ref{fig5}(b), right column. 
As a results, the electron density and the envelope  function of the ground state are localized in the center of the NW and exhibit a circular symmetry.
The charge distribution is hardly modulated by the potential applied to the gate, and only slightly shifted upward or downward, depending on the sign of the gate potential [Fig.~\ref{fig5}(a)(c)], and the SOC coefficient changes sign accordingly. Moreover, since the gate is located at the bottom of the structure, positive voltages are slightly more effective in shifting the envelope function downward, see Fig.\ref{fig5}(a,c), hence the slight asymmetry between positive and negative voltages shown in Fig.~\ref{fig4}(c). 

\begin{figure}[!ht]
	\begin{center}
		\includegraphics[scale=.4]{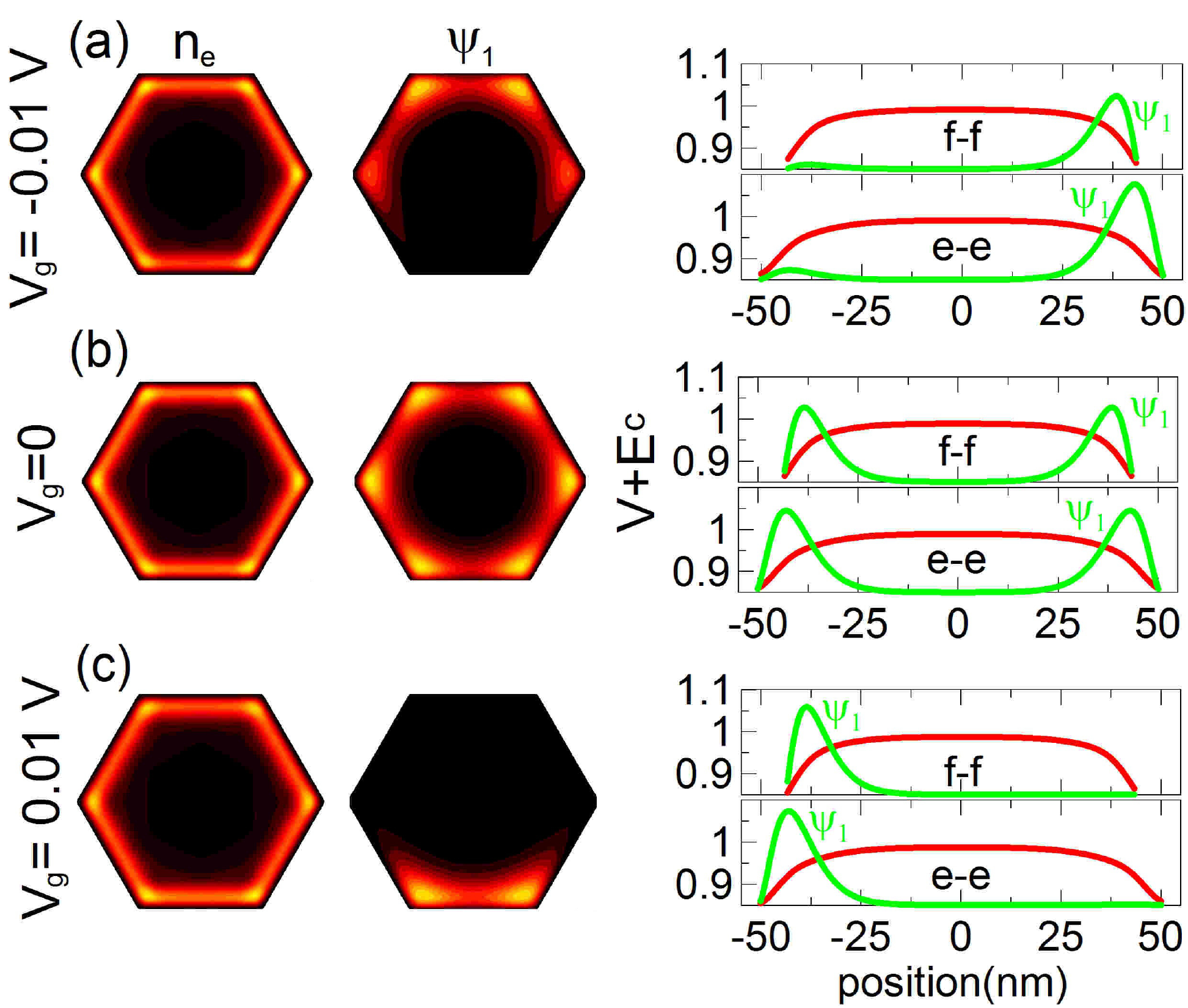}
		\caption{Same as Fig.~\ref{fig5} for gate voltage (a) $V_g=-0.01$~V, (b) $V_g=0$ and (c) $V_g=0.01$~V. Calculations performed with  $n_e=3\times10^9$~cm$^{-1}$.}
		\label{fig6}
	\end{center}
\end{figure}

At the high concentration regime the electron-electron interaction dominates and total energy is minimized by reducing repulsive Coulomb energy, at the expense of localization energy. Accordingly, electrons move outwards and accumulate near the facets. At sufficiently high electron concentration charge localize in quasi-1D channels at the edges,\cite{Bertoni2011} a minor part of the charge sits at the facets, while the core of the wire is totally depleted, as shown in Fig.~\ref{fig6}(b). The strong localization of the ground state at the six edges of the hexagon explains the shooting of the SOC around $V_g=0$. 
Indeed, since localization in the core (hence tunneling energy between oppositely localized states) vanish, symmetric edge localization is easily destroyed by any slight asymmetry introduced by the gate potential. 
A similar, more common situation, occurs in coupled symmetric quantum wells~\cite{Calsaverini2011} when the symmetric and antisymmetric states are nearly degenerate. 
As presented in Fig.~\ref{fig6}(a,c), any slight positive or negative voltage applied to the gate results in the localization of the ground state in the two lower or upper edges, respectively. Accordingly, the SOC coefficient abruptly changes from zero and almost saturates in a narrow range around $V_g=0$. 

In other words, the SOC coefficient is a sensitive probe of the complex localization of the charge density in different regimes. To make this aspect more quantitative, we define a SOC susceptibility $\chi= d \alpha ^{11}_x / dV_g |_{V_g=0}$, i.e., the slope of $\alpha^{11}_x$ at zero gate voltage. Its dependence on the charge density, shown in Fig.~\ref{fig4}(d), is clearly non linear and correlated to the strength of the Coulomb interaction. Note that, although $\chi$ grows with charge density, there is no sign of a critical behavior in our mean-field calculations.

\begin{figure}[!ht]
	\begin{center}
		\includegraphics[scale=.35]{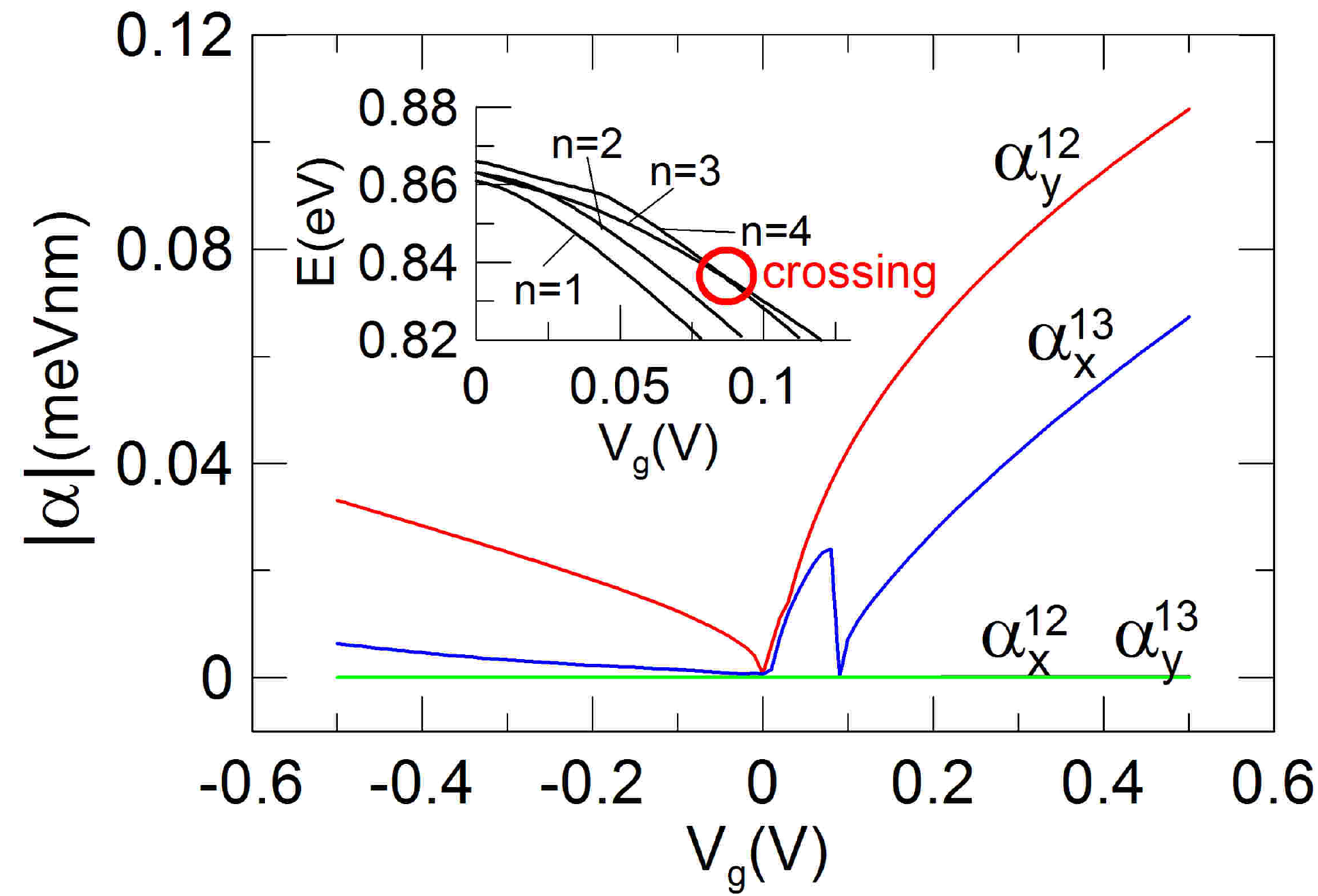}
		\caption{Absolute value of the inter-subband SOC couplings $\alpha^{12}_{x(y)}$ and $\alpha^{13}_{x(y)}$ as a function of the gate voltage $V_g$. 
			Inset shows $E_n(V_g)$ for the four lowest electronic states. Results for $n_e=10^7$~cm$^{-3}$.}
		\label{fig7}
	\end{center}
\end{figure}

\begin{figure}[!ht]
	\begin{center}
		\includegraphics[scale=.4]{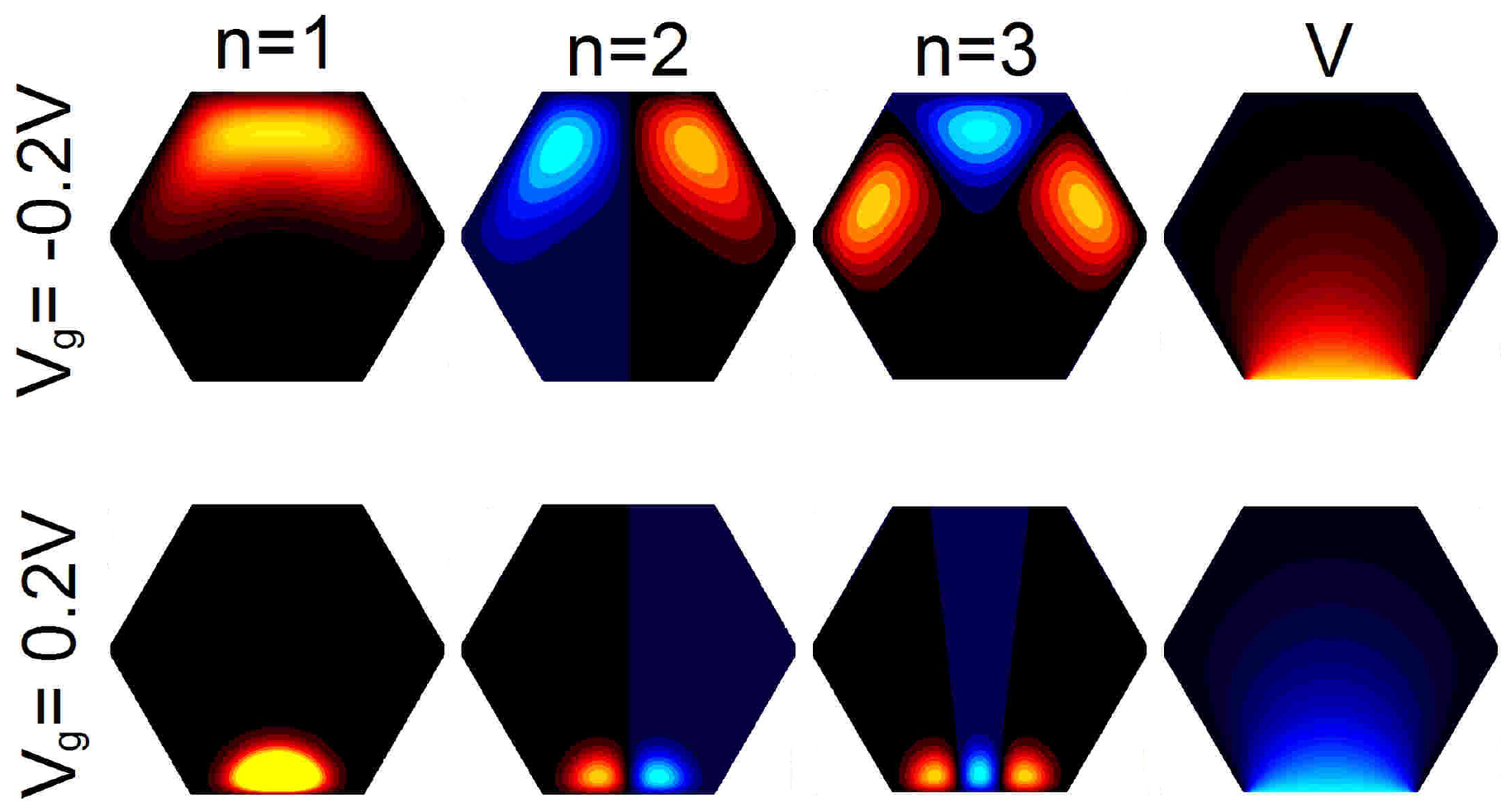}
		\caption{Envelope  functions of the three lowest electronic states together with the self-consistent potential $V(x,y)$ for $V_g=-0.2$~V and $V_g=0.2$~V.}
		\label{fig8}
	\end{center}
\end{figure}

Off-diagonal terms in Eq.~(\ref{eq:a}) represent spin-flip processes combined to inter-subband scattering. Such inter-subband SOC has been related to intriguing physical phenomena, such as unusual Zitterbewegung,\cite{Bernardes2007} intrinsic spin Hall effect in symmetric quantum well\cite{Hernandez2013} and spin filtering devices.~\cite{Wojcik2016,Wojcik2017} Below we analyze inter-subband SOC between the ground state and the two lowest excited states.

Figure~\ref{fig7} shows $|\alpha^{12}_{x(y)}|$ and $|\alpha^{13}_{x(y)}|$ as a function of the gate voltage $V_g$ in the low electron concentration regime, $n_e=10^7$~cm$^{-3}$. 
In the whole range of $V_g$, these coefficients remain almost one order of magnitude smaller than the the intra-subband coefficient $\alpha^{11}_{x}$. The  discontinuity of $\alpha^{13}_{x}$ in Fig.~\ref{fig7} is caused by the crossing of subbands $n=3$ and $n=4$ at $V_g \approx 0.08$~V, as shown in the inset.
Note that $\alpha^{12}_{x}=\alpha^{13}_{y}=0$ due to the symmetry of the envelope functions and the self-consistent potential $V(x,y)$, as illustrated in Fig.~\ref{fig8}, where the first three states and the corresponding potential are reported, for two opposite gate voltages.

\begin{figure}[!ht]
	\begin{center}
		\includegraphics[scale=.35]{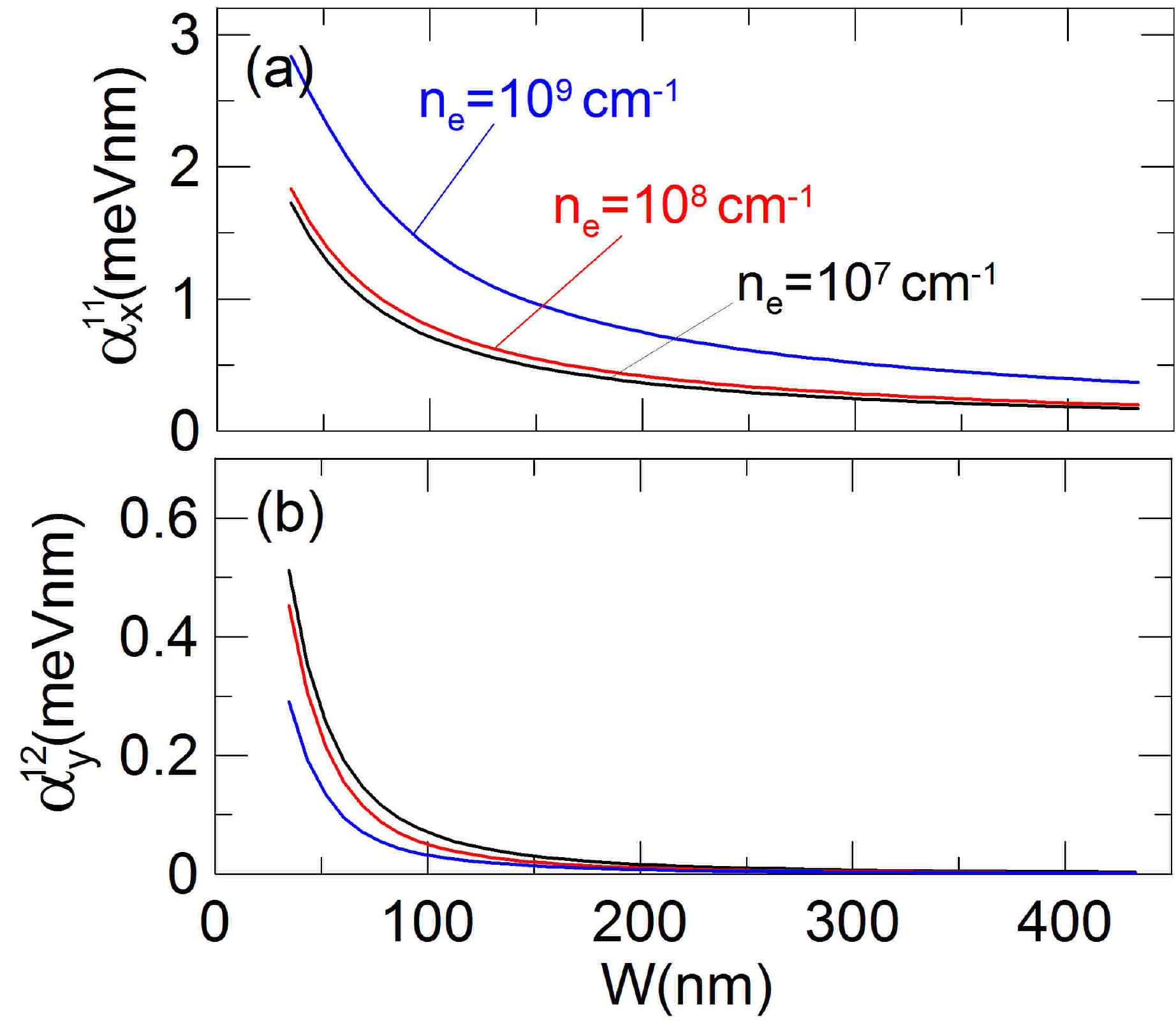}
		\caption{(a) The intra-subband $\alpha^{11}_{x}$ and (b) inter-subband $\alpha^{12}_{y}$ couplings as a function of the NW width $W$, for different electron concentrations, as indicated, at $V_g=0.4$~V.}
		\label{fig9}
	\end{center}
\end{figure}

Finally, in Fig.~\ref{fig9} we show the behavior of intra- and inter-subband  SOC couplings with varying NW width, showing monotonous decrease with increasing width. As expected, in wide NWs SOC tends to zero.

\subsection{InSb}
\label{sec:InSb}

Indium antimonide (InSb) is a strong SOC material, due to its low energy gap and small conduction electron mass, which makes this semiconductor the preferred host material for spintronic applications and topological quantum computing. 

\begin{figure}[!ht]
	\begin{center}
		\includegraphics[scale=.42]{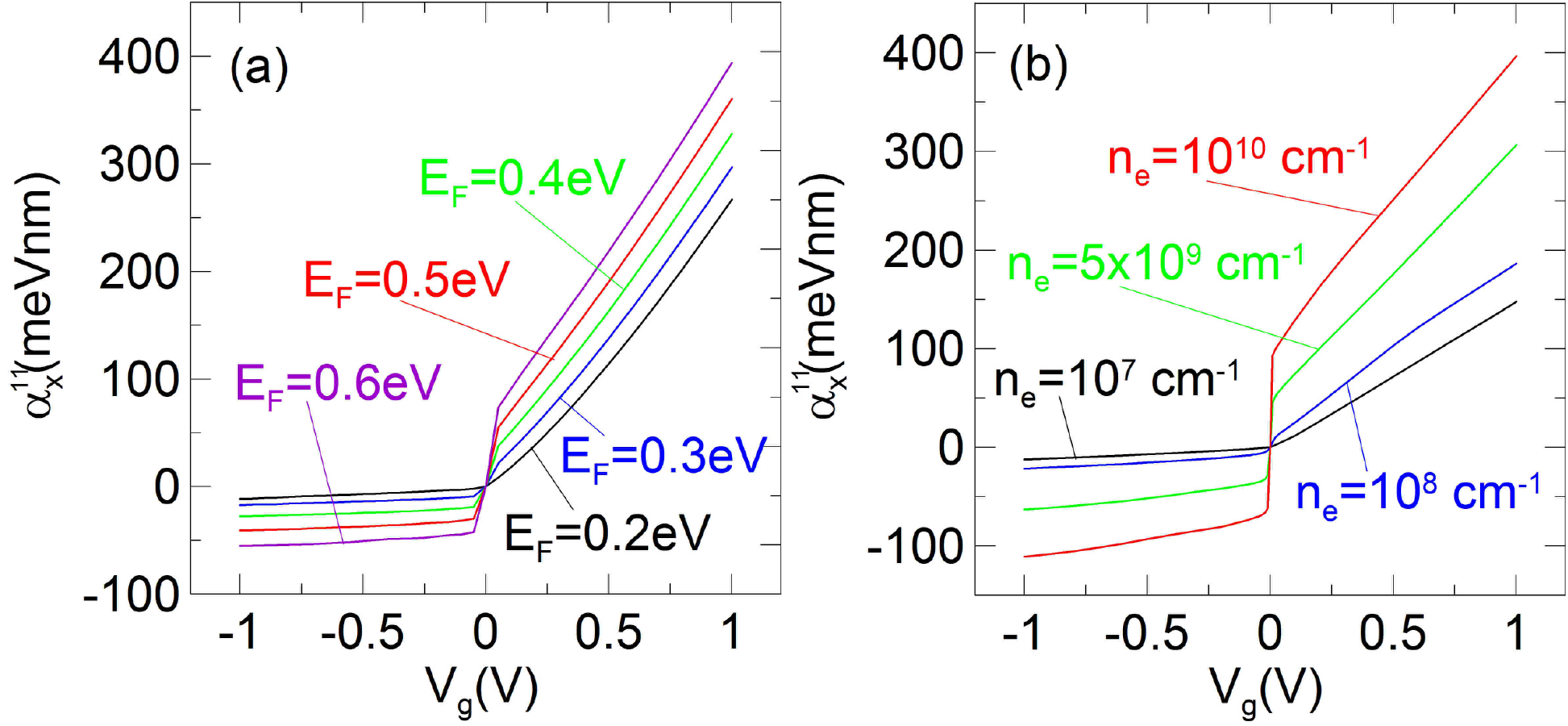}
		\caption{$\alpha^{11} _x$ vs $V_g$ calculated by the assumption of (a) the constant chemical potential and (b) the constant electron density. Results for $W=87$~nm and the 'ideal' bottom gate configuration.}
		\label{fig10}
	\end{center}
\end{figure}

In this section, we investigate SOC in InSb NWs in the context of recent experiments\cite{vanWeperen2015,Kammhuber2017} reporting extremely high value of the SOC coefficients. 
Calculations shown below have been carried out for the following material parameters:\cite{Vurgaftman2001} $E_0=0.235$~eV, $\Delta_0=0.81$~eV, $m^*=\overline{m}^*=0.01359$, $E_P=2m_0P^2 /\hbar^2=23.3$~eV and dielectric constant $\epsilon=16.8$. 
As in the previous case, we consider a temperature $T=4.2$~K.

To compare the SOC in InSb and GaAs NWs, in Fig.~\ref{fig10} we present  $\alpha^{11} _x$ for the bottom gate configuration and calculation models used in the previous subsection, i.e. $W=87$~nm and the 'ideal' gate configuration. 
The  behavior is qualitatively similar to GaAs NWs [see Fig.~\ref{fig4}(a,b)] but SOC coefficients are two orders of magnitudes larger in InSb NWs. 
We next investigate two specific configurations to compare explicitly with recently reported experimental setups.

\subsubsection{Comparison with Ref.~[\onlinecite{vanWeperen2015}]}

\begin{figure}[!ht]
	\begin{center}
		\includegraphics[scale=.42]{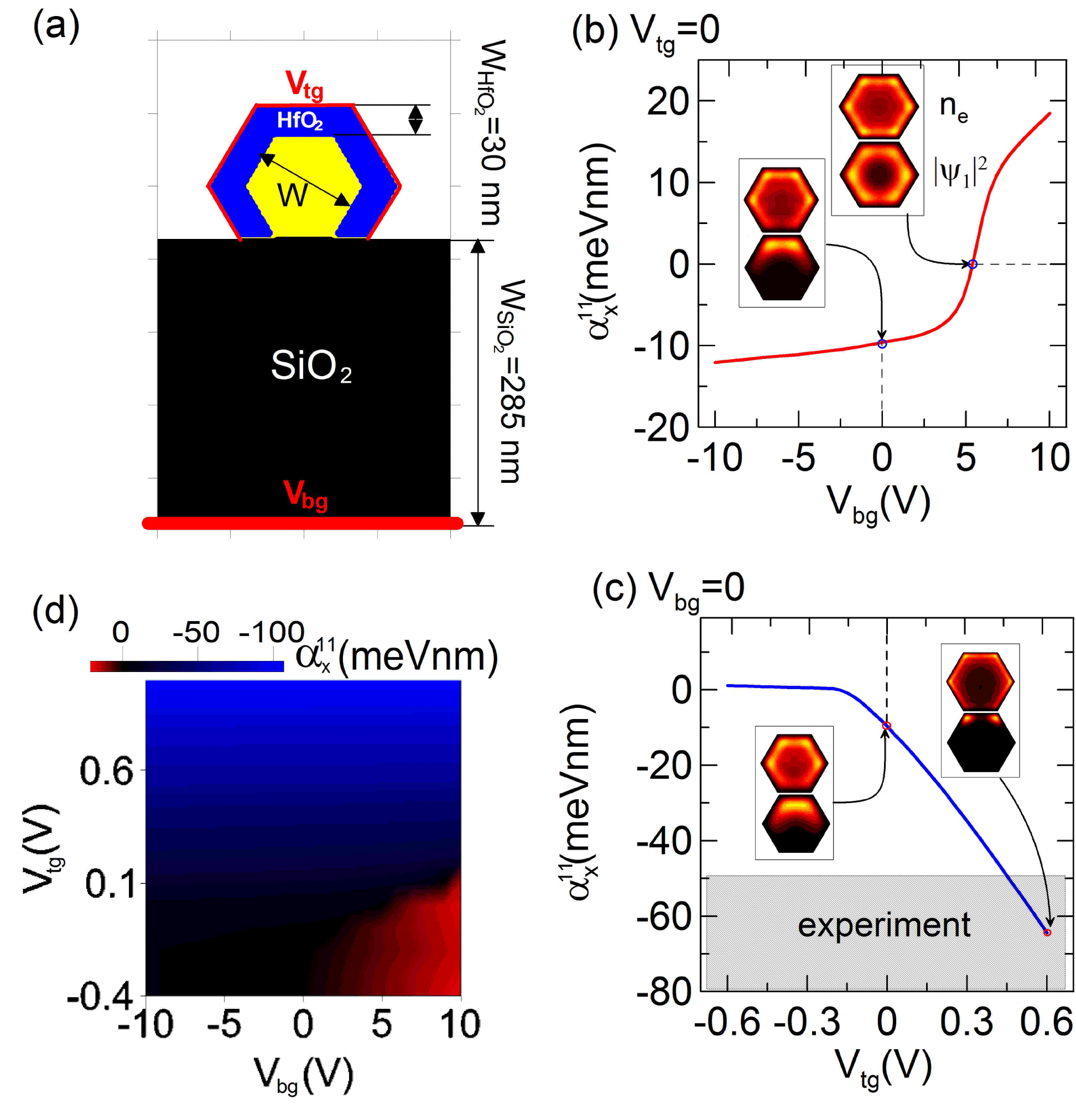}
		\caption{(a) Schematic illustration of the simulated device. Two gates are connected to the wire through dielectric layers, as indicated, held at voltages $V_{bg}$ and $V_{tg}$. (b) $\alpha^{11} _x$ as a function of the bottom gate voltage
			$V_{bg}$, with $V_{tg}=0$. (c) $\alpha^{11} _x$ as a function of the top gate voltage $V_{tg}$, with $V_{bg}=0$. Insets in panels (b) and (c)
			show the electron concentration $n_e(x,y)$ (top) and square of the envelope  functions of the ground state $|\psi_1(x,y)|^2$ (bottom) at selected gate voltages indicated by arrows. In panel (c) the range of measured SOC coefficient\cite{vanWeperen2015} is marked by the
			gray area. (d) Map of $\alpha^{11} _x$ as a function of both the gate voltages $V_{tg}$ and $V_{bg}$. Results for
			$\mu=0.35$~eV. }
		\label{fig11}
	\end{center}
\end{figure}

In Ref.~\onlinecite{vanWeperen2015} the authors used magnetoconductance measurements in dual-gated InSb NW devices, with a theoretical analysis of weak antilocalization to extract the SOC coefficients.
They obtained SOC coefficients as large as $50-100$~meVnm. In the measurements, the conductance of the NW was controlled by a back gate, separated from the wire by a $285$~nm thick  SiO$_2$ layer, and a $\Omega$-shaped gate, separated by a HfO$_2$ layer $30$~nm thick. 
The schematic illustration of the experimental setup is shown in Fig.~\ref{fig11}(a).

As in experiments, we consider the NW with a width $W=100$~nm and sweep the gate voltages $V_{bg}=[-10~\text{V},10~\text{V}]$ and $V_{tg}=[-0.6~\text{V},0.6~\text{V}]$. 
Simulations have been carried out in the $\mu$-constant model, $\mu$ being the only free parameter of the calculations. 
Its value has been determined on the basis of the conductance measurements shown in Fig.~3 of Ref.~\onlinecite{vanWeperen2015}, which indirectly show the occupation of subsequent electronic states in the NW while changing both gate voltages  $V_{bg}$, $V_{tg}$.
Comparing the occupation map from our simulations with the experimental conductance map, we estimate $\mu=0.35$~eV. For this value, and $V_{bg}=V_{tg}=0$, $N=15$ subbands are occupied, in agreement with the estimated value reported in Ref.~\onlinecite{vanWeperen2015}.

Figure \ref{fig11}(b) shows $\alpha^{11} _x$ as a function of the bottom gate voltage, $V_{bg}$, with $V_{tg}=0$ ($\alpha^{11} _y$ is zero by symmetry for this gate configuration). 
Note that, due to the geometrical asymmetry related to the different position of the gates with respect to the NW, $\alpha^{11} _x \neq 0$ even at $V_{bg}=V_{tg}=0$. The ''symmetry point'' $\alpha^{11} _x = 0$ is obtained at $V_{bg}\approx 5.4$~V, compensating for the electrostatic asymmetry caused by the experimental geometry. Around this value, $\alpha^{11} _x$ rapidly changes sign. Moreover, due to the weak coupling of this gate to the NW, $\alpha^{11} _x$ varies only slightly in the considered voltage range. 

The situation is different sweeping the voltage of the strongly coupled top gate. In this case $\alpha^{11} _x$ grows by almost two orders of magnitude with increasing $V_{tg}$, as shown in Fig.~\ref{fig11}(c). The asymptotically vanishing of $\alpha^{11}_x$ at large, negative $V_{tg}$ is due to full depletion of the NW. Interestingly, for $V_{tg}$ in the (positive) range $[0.45-0.6]~V$, SOC achieves values comparable to experiments, $|\alpha^{11}_x|\approx 50-100$~meVnm. 
Since for $V_{tg}=0.6$~V up to $N=45$ subbands are occupied, the large value of $\alpha^{11} _x$ is mainly determined by the electron-electron
interaction, through the localization mechanism already described for GaAs NWs. For completeness, the full map $\alpha^{11}
_x (V_{bg},V_{tg})$ is presented in Fig.~\ref{fig11}(d).

\begin{figure}[!ht]
	\begin{center}
		\includegraphics[scale=.35]{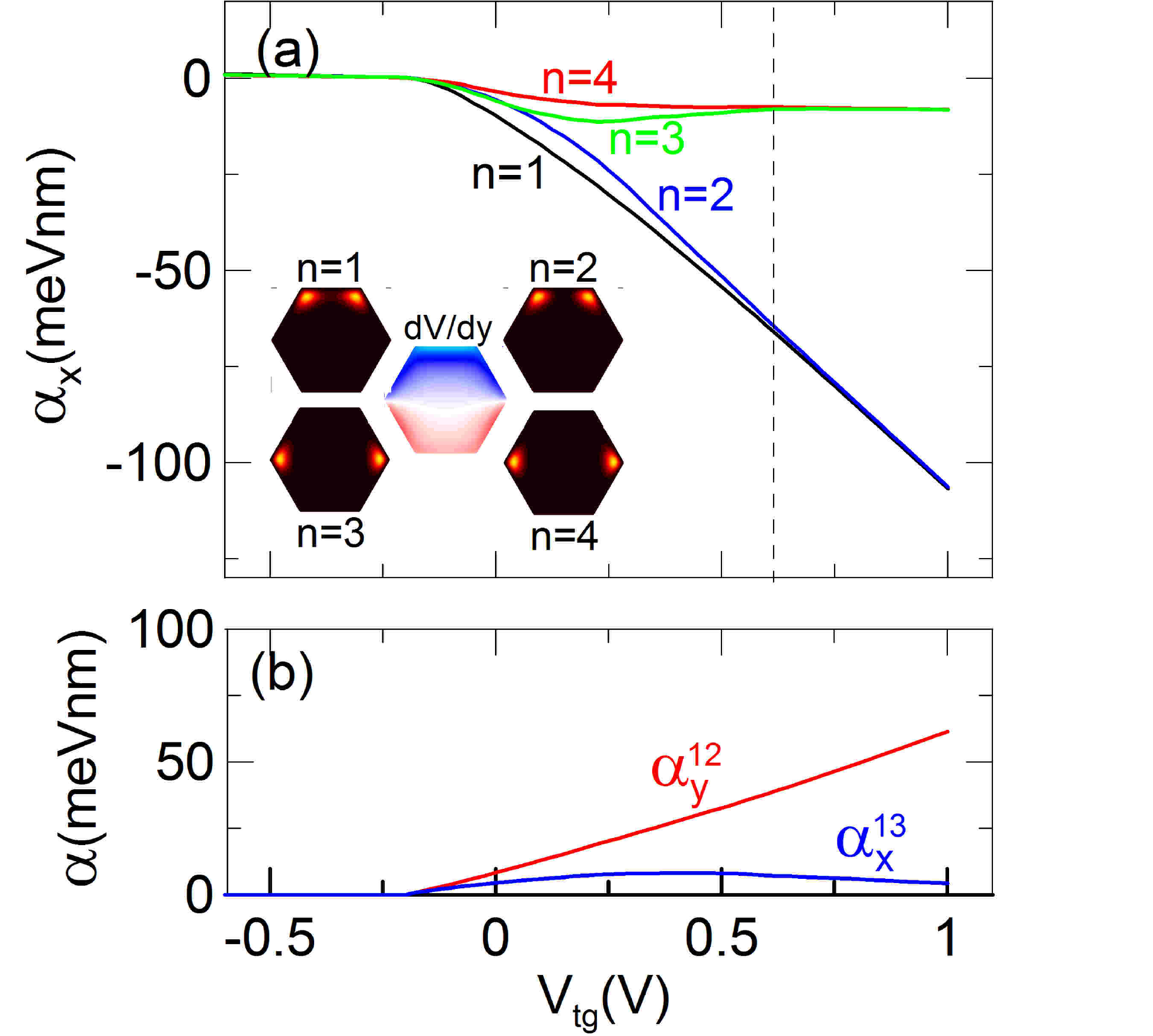}
		\caption{(a) SOC couplings $\alpha^{nn} _x$ as a function of the top gate voltage $V_{tg}$ calculated for the four lowest subbands ($V_{bg}=0$). The corresponding $\left|\psi_n\right|^2 $ are shown in the inset for $V_{tg}=0.6$~V (vertical dashed line), together with the $y$-component of the electric field. (b) Inter-subband SOC couplings $\alpha^{12} _y$ and $\alpha^{13} _x$  as a function of the top gate voltage $V_{tg}$.}
		\label{fig12}
	\end{center}
\end{figure}

For experimental setups characterized by a strong geometrical asymmetry, $\alpha^{nn} _x$ may be different for different subbands. In Fig.~\ref{fig12}(a) we present $\alpha^{nn} _x$ as a function of  $V_{tg}$ for the four lowest electronic states. For $n=1,2$ $\alpha _x$ decreases with increasing $V_{tg}$, taking the absolute value $50-100$~meVnm in agreement with the experiment. Although near to $V_{tg}=0$ the curves are different, for large positive voltages, when the electron-electron interaction is dominant, the curves approach each other. For states $n=3,4$, on the other hand, $\alpha _x$ quickly saturates at $\alpha _x \approx - 10$~meVnm and it is almost unaffected by large positive gate voltages. This behavior can be traced to the different localization of the envelope functions for different subbbands, shown in the inset of Fig.~\ref{fig12} for $V_{tg}=0.6$~V. Indeed, the envelope functions of states $n=3,4$ are localized on opposite corners, in regions where the gradient of the potential changes 
sign, 
resulting in a strong reduction of the SOC constant for these subbands [see. Eq.(\ref{eq:a})].\\

From the experimental point of view, it is interesting to evaluate the inter-subband SOC, shown in  Fig.~\ref{fig12}(b). For the present NW diameter, which is quite large, the gap between the lowest two subbands is $\Delta E_{12}\approx 4$~meV at $V_{tg}=0$ and decreases down to $\approx 2$~meV at $V_{tg}=0.6$~V, when both  states are strongly localized in the two top corners and differ only by the parity. As a results, $\alpha ^{12} _{y}$ reaches values close to that of the intra-subband coefficient $\alpha^{11}_x$ which means that the spin dynamic of the electron in the ground state is determined  equally by both the intra- and inter-subband SOC.

\subsubsection{Comparison with Ref.~\onlinecite{Kammhuber2017}}

\begin{figure}[!ht]
	\begin{center}
		\includegraphics[scale=.42]{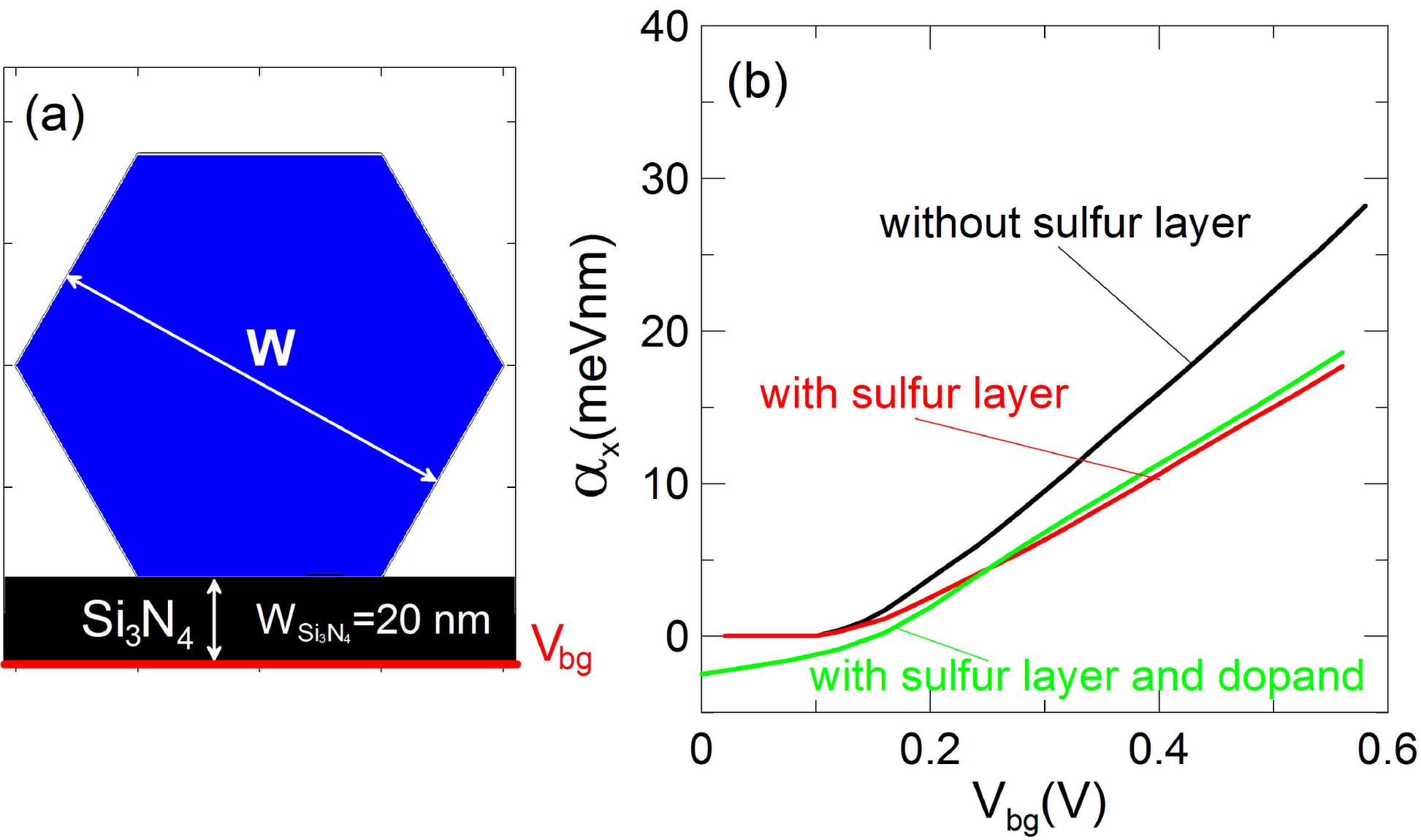}
		\caption{(a) Schematic illustration of the experimental setup in Ref.~\onlinecite{Kammhuber2017}. The conductance of the NW is controlled by the bottom gate $V_{bg}$ attached to the wire by the $20$~nm thick Si$_3$N$_4$ layer. (b) $\alpha^{11} _x$ as a function of the bottom gate $V_{bg}$ for the NW without the sulfur layer (black curve), with the sulfur layer (red curve), and with the dopant concentration included (green curve).}
		\label{fig13}
	\end{center}
\end{figure}

In Ref.~\onlinecite{Kammhuber2017} the authors reported the conductance measurements through the helical gap in InSb NWs. 
Analysis of the experimental data, taken at different magnetic field orientations,using a single-electron model, led to an extremely large SO energy $E_{SO}=6.5$~meV, corresponding to $\alpha_R\approx 270$~meVnm. 
The conductance of the NW was controlled by a bottom gate attached to the wire through a $20$~nm thick Si$_3$N$_4$ layer, while all other facets were electrostatically free. 
In calculations the Neumann boundary conditions were applied. 
The schematic illustration of the sample used in the experiment is reported in Fig.~\ref{fig13}(a).

To investigate the device described in Ref.~\onlinecite{Kammhuber2017}, we simulated a NW of width $W=100$~nm and gate voltage in the range $V_{bg}=[0~\text{V},0.6~\text{V}]$. 
Simulations have been carried out in the $\mu$-constant model, and $\mu$ was chosen on the basis of the conductance measurements in Ref.~[\onlinecite{Kammhuber2017}], reporting the first conductance step at $V_g\approx0.1$~V. 
In our simulations such an occupation is realized with $\mu=40$~meV.

Figure \ref{fig13}(b) shows $\alpha^{11} _x$ as a function of the bottom gate $V_{bg}$. The calculated value of the SOC coefficient is about nine times lower than reported in the experiment. In an attempt to explain this discrepancy we referred to details of the nanofabrication.\cite{Zhang2016} The precise procedure for the contact deposition includes etching of the native oxide at the InSb NW using sulfur-based solution. Inclusion sulfur at the InSb surface may produce a variable donor concentration up to $7.5 \times 10^{18}$~cm$^{-3}$,\cite{Wieder1968} which results in band bending with electron accumulation near the surface. Accordingly, in our calculations we included a $5$~nm thick sulfur layer at the InSb NW surface, and considered two cases: without dopants and with a dopant concentration $n_d=10^{17}$~cm$^{-3}$. As shown in Fig.~\ref{fig1}(b), the presence of the sulfur layer decreases the SOC coefficient, due to the low dielectric constant and the reduction of the electric field in the NW. 
Even inclusion of the dopants, which bends the conduction band at the interfaces, does not change this behavior qualitatively, leaving our results well below the experimentally measured SOC constant.

Therefore, this discrepancy remains unexplained. Note that such a large SOC constant has been reported only in one experiment so far,~\cite{Kammhuber2017} fitting the helical state conductance measurements to a single-band model which includes neither the orbital effects nor the inter-subband coupling. 
Both these effect may increase the effective SOC in the ground state and the use of the simple single band theory to extract the $\alpha$ value can lead to overestimation of this parameter.

\section{Summary}
\label{sec:summary}

We have formulated a multi-band $\vec{k}\cdot\vec{p}$ theory of SOC in NW-based devices and investigated the behavior of Rashba SOC in GaAs- and InSb-based devices. 
The strength of the SOC coefficients are determined by band parameters and external potentials. 
In the absence of any external potentials, the charge density shares the symmetry of the structure, hence SOC coefficients vanish. 
External gates, breaking the symmetry, can tailor SOC.
The tenability of the SOC coefficients, however, strongly depend on size and doping. 
We show, for example, that in the high carrier density regime SOC has a very large susceptibility.

In light of our simulations, we analyzed quantitatively recent experiments with InSb nanowires. Good agreement is found with SOC reported in Phys.~Rev.~B \textbf{91}, 201413(R) (2015), but not with the much larger values measured in Nat~Commun., \textbf{8}, 478 (2017). We argue that possible origins of this discrepancy lies in the model used to extract the parameter, which entails a single-particle, single band model. Our calculations, on the contrary, show that electron-electron interaction plays a dominant role and inter-subband contributions are substantial in the investigated samples. 

\section{Acknowledgement}
This work was partially financed (supported) by the Faculty of Physics and Applied Computer Science AGH UST dean grant for PhD students and young researchers within subsidy of Ministry of Science and Higher Education and in part by PL-Grid Infrastructure.


%

\end{document}